\documentclass{PoS}
\usepackage{amsmath}
\usepackage{graphicx}
\usepackage[square,numbers]{natbib}
\usepackage{subfig}

\usepackage{epsfig} 
\usepackage{ifpdf} 
\usepackage{fancyvrb}
\usepackage{txfonts}
\usepackage{xcolor}
\usepackage{soul}
\usepackage{mathtools}

\title{FACT -- Multi-wavelength analysis of more than 30 flares of Mrk 421}

\ShortTitle{Multi-wavelength analysis of more than 30 flares of Mrk 421}

\author{
V.\,Sliusar$\,^1$,  A.\,Arbet-Engels$\,^2$,  D.\,Baack$\,^3$,  
M.\,Balbo$\,^1$,
M.\,Beck$\,^{2,a}$,  A.\,Biland$\,^2$,   M.\,Blank$\,^4$,  
T.\,Bretz$\,^{2,a}$,
K.\,Bruegge$\,^3$,  M.\,Bulinski$\,^3$,  J.\,Buss$\,^3$,  
M.\,Doerr$\,^4$,
D.\,Dorner$\,^4$,  D.\,Elsaesser$\,^3$,  D.\,Hildebrand$\,^2$,  
R.\,Iotov$\,^4$,
M.\,Klinger$\,^{2,a}$, K.\,Mannheim$\,^4$, S.A.\,Mueller$\,^2$, 
D.\,Neise$\,^2$,
M.\,Noethe$\,^3$,  A.\,Paravac$\,^4$,  
W.\,Rhode$\,^3$,
B.\,Schleicher$\,^4$,  K.\,Sedlaczek$\,^3$,  A.\,Shukla$\,^4$,  
L.\,Tani$\,^2$,
F.\,Theissen$\,^{2,a}$,
\speaker{R.\,Walter}$\,^1$ and
E.\,von Willert$\,^4$

(the~FACT~Collaboration\footnote{for collaboration list see 
PoS(ICRC2019)1177} \,)\\

    {$^1$}University of Geneva, Department of Astronomy,
     Chemin d'Ecogia 16,  1290 Versoix, Switzerland\\
    {$^2$}ETH Zurich, Institute for Particle Physics and Astrophysics,
     Otto-Stern-Weg 5, 8093 Zurich, Switzerland\\
    {$^3$}TU Dortmund, Experimental Physics 5,
     Otto-Hahn-Str. 4, 44221 Dortmund, Germany\\
    {$^4$}Universit\"at W\"urzburg, Institute for Theoretical Physics and 
Astrophysics,
     Emil-Fischer-Str. 31, 97074 W\"urzburg, Germany\\
    {$^a$}also at RWTH Aachen University, Physics Institute III A, 52074 
Aachen, Germany\\
   E-mails: \email{vitalii.sliusar@unige.ch}, 
\email{roland.walter@unige.ch} \medskip\\
}

\abstract{
Mrk 421 is a high-synchrotron-peaked blazar featuring bright and persistent GeV and TeV
emission. We use multi-wavelength light curves of Mrk 421 spanning 5.5 years with FACT
(TeV) and Fermi LAT (GeV) in the gamma rays, Swift BAT, Swift XRT and MAXI in the X-rays,
together with optical and radio data and investigate the physical processes driving the emission
and variability. Observations by FACT are continuous and not triggered, so the source was found
in a wide range of flux states and more than 30 flares were identified from X-rays to TeV. The light curves in  TeV and X-rays feature very similar flares with rise and decay times of a few days and zero lag,
characteristic for electron processes. At least two parameters per flare, the amplitude and the
cut-off energy, are required to explain the observed variability. In addition, the GeV light curve leads and
is strongly correlated with the optical and radio light curves as expected from SSC emitting shock propagating in a conical jet.
}

\FullConference{36th International Cosmic Ray Conference -ICRC2019-\\
		July 24th - August 1st, 2019\\
		Madison, WI, U.S.A.}

\begin{document}

 \makeatletter
 \setbox\@firstaubox\hbox{\small V.~Sliusar}
 \makeatother

\section{Introduction\label{sec:introduction}}

Mrk\,421 is one of the brightest and closest high-energy-peaked blazars ($z = 0.031$). It features a bright and persistent GeV and TeV emission with frequent flaring activities. Its average spectral energy distribution has been fitted with different models: e.g.: one zone leptonic synchrotron self-Compton model (SSC) \citep{abdo_2011ApJ...736..131A}, hadronic model where the accelerated protons cool through synchrotron emission \citep{2015MNRAS.448..910C} or interact with the leptonic synchrotron photons to create cascade of pions and muons, decaying in $\gamma$-rays and neutrinos \citep{2001APh....15..121M}. Fast variations suggest, that either a shock moves relativistically within the jet or the emission region is much smaller than the gravitational radius \citep{2014A&A...563A..91A}, and is driven by the interactions between stars/clouds and the jet or by magnetic reconnections.

To the date, Mrk\,421 was a target of multiple multi-wavelength campaigns  \citep[e.g.][]{2004ApJ...601..759T,2015A&A...576A.126A}. During the 2009 campaign, Mrk\,421 was simultaneously observed from the radio to the TeV band for 4.5 months in the absence of a strong flare \citep{2015A&A...576A.126A}. The fractional variability revealed that most variability lies in the X-ray ($F_{var}=0.5$) and TeV ($F_{var}=0.3$) bands. A harder-when-brighter behaviour was discovered in the X-rays. The smallest variations were observed in the radio. A positive cross-correlation was found between X-rays and TeVs at a maximum lag of $\sim$5 days. No strong correlations between optical/UV and X-rays were found.

In this paper, we report results from a MWL analysis using 5.5 years of the FACT monitoring in the TeV band, which is ongoing for more than 7 years \citep{2013arXiv1311.0478D}. The observations were not triggered, therefore unbiased, and as regular as possible taking into account observing conditions constraints. We also used continuous radio, optical, ultraviolet, X-ray, GeV and TeV light curves obtained quasi-simultaneously with FACT with the aim to investigate various emission models.

\section{Multi-wavelength data\label{sec:data}}

Data from nine different instruments was used to build the light curves spanning between December 14, 2012 and April 18, 2018. We performed the cross-correlation, auto-correlation, Bayesian Block and fractional variability analysis using this dataset to investigate the physical processes responsible for the emission in all bands. During the considered period,  Mrk\,421 was observed at various flux states in all bands. Flares observed from the X-rays to the TeV are narrow enough to be identified individually (see table~\ref{tab:flares}). At longer wavelength the flares become wider and eventually overlap. 

The TeV data were obtained by the First G-APD Cherenkov Telescope (FACT), which is a 3.8\,m imaging air Cherenkov telescope located at La Palma \citep{2013JInst...8P6008A}. Since the start of data taking in 2011, the system gradually was automatized, resulting in remote operation since late July 2012 and fully robotic operation since 2017. This allows to perform long-term
regular observations of bright TeV sources with high cadence. Due to the use of a SiPM camera and a state-of-the-art feedback system, the telescope can operate in bright ambient light conditions \citep{2013arXiv1311.0478D}. A detailed description of events reconstruction and analysis is presented in \citep{2017ICRC...35..779H}. The quality checks and background suppression techniques are described in \citep{2017ICRC...35..612M}, \citep{2019arXiv190203875B} and \citep{2005ICRC....5..215R}. The energy threshold based on simulated data is determined to be 775\,GeV for sources with Crab-Nebula-like spectrum and $\sim$860\,GeV for Mrk\,421 (harder spectrum).

The Mrk\,421 data in GeV $\gamma$-rays with energy 100 MeV $<$ E $<$ 300\,GeV was obtained by the Fermi Large Area Telescope (LAT)\citep{2009ApJ...697.1071A}. The PASS8 pipeline and Fermi Science Tool v10r0p5 package were used to process the data. The fitting model included sources from the LAT 4-year Point Source Catalogue.

X-ray observations in the 15-50\,keV band were performed by Swift/BAT. The reduction pipeline is based on the BAT analysis software \texttt{HEASOFT} version 6.13 \cite{2013ApJS..207...19B}. The daily light curve was built using observational data from 29344 orbital periods or almost 5.5 years. 

X-ray observations in the 0.2-10\,keV band were performed by the Swift/XRT X-ray telescope \citep{2005SSRv..120..165B}. The light curve was obtained from the on-line Swift-XRT products generation tool \footnote{http://www.swift.ac.uk/user\_objects/}, which uses \texttt{HEASOFT} software version 6.22.

X-ray observation from 2 to 20\,keV were performed by MAXI instrument \citep{2009PASJ...61..999M}. The light curve for Mrk\,421 from MAXI is publicly available \footnote{http://maxi.riken.jp/star\_data/J1104+382/J1104+382.html}.

The UV observations were performed by Swift/UVOT telescope in three bands UVW1, UVM2 and UVW2 \citep{roming_2005SSRv..120...95R}. The data were reduced using on-, off-method using the HEASOFT package version 6.24 along with UVOT CALDB version 20170922.

The V-band data from 1.54\,m Kuiper Telescope on Mountain Bigelow and the 2.3\,m Bok Telescope on Kitt Peak were used for the monitoring \citep{2009arXiv0912.3621S}. Our dataset includes data from Cycle 5 to 10. The light curves 
are available publicly\footnote{http://james.as.arizona.edu/$\sim$psmith/Fermi/DATA/photdata.html}.

Mrk\,421 observations are regularly performed in the radio by the OVRO 40 meter radio telescope. Observations were performed twice per week at 15 GHz. The data are available from the telescope archive publicly\footnote{http://www.astro.caltech.edu/ovroblazars/}.

\section{Light curves timing and correlation analysis}
Using multi-wavelength light curves, we calculated values of fractional variability, lags and correlations between different bands, found best-fit parameters of the GeV to radio response profile.  

\subsection{Fractional variability}
We performed the fractional variability analysis of all the light curves of Mrk\,421 as proposed in \citep{Vaughan_2003MNRAS.345.1271V}. The uncertainties were estimated following the \cite{Poutanen_2008MNRAS.389.1427P}. Using the 5.5 years of multi-wavelength observations, we found that the lowest $F_{var}=0.15$ is in radio and the highest $\sim$1.33 in X-rays (Swift/BAT). Then the fractional variability drops to $\sim$0.34 for Fermi LAT and increases to $\sim$0.92 in TeV making two distinctive humps in the $F_{var}$ dependency of the frequency. Previous studies of the fractional variability were reported in \citep{2015A&A...576A.126A} and \citep{aleksic_2015A&A...578A..22A}, while having the same two hump behaviour, the amplitude was three time lower, which likely can be explained by much shorter light curves being used.

Two maxima of the fractional variability of Mrk\,421 indicate that the high energy portions of the two emission components are more variable than the low energy ones. This suggests that the cutoff energies of both components are the primary source of the variability. Similarly, we verified that the simultaneous flux for the TeV and X-ray is well linked, underlying that a single parameter is driving these variations.
 

\subsection{Correlations}
To investigate the connection between the emission in different bands, we calculate the cross-correlation between all the light curves. We use DCF \citep{Peterson_1998PASP..110..660P} to estimate the correlation between irregularly sampled data. We adopted 1 day bins for most DCFs due to long light curves being used (except when the data were more sparse). Uncertainties in DCF are in most cased underestimated, so we adopted Monte Carlo simulations to calculate the lag probability. The lag uncertainty corresponds to the standard deviation of the distribution of the lags obtained for the random subsets.

To estimate the variability timescales, we calculate the discrete auto-correlation functions from radio to TeV. The uncertainties are estimated using Monte Carlo simulations similarly to the case of DCFs. We find that variability time scale observed in the TeV and X-ray bands are short, $\sim$3 days, which is consistent with the models where emission is due to fast relativistic electron cooling. The same quantity for the GeV Fermi LAT light curve is even smaller ($\lesssim1$ day), which might be an evidence of different parameters driving the variability in that band.

\begin{figure}
  \centering
       \includegraphics[width=0.83\columnwidth]{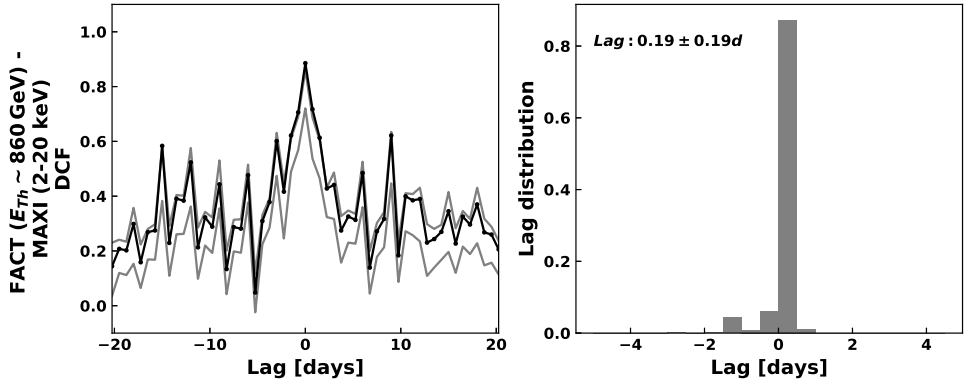}
       \caption{DCF cross-correlations of TeV (FACT) and X-rays (MAXI) light curves. One day binning was used. Left: DCF values as a function of lag. Gray lines denote the 1$\sigma$ uncertainties. Right: lag distribution corresponding to the maximum DCF value.}
  \label{fig:dcf_all}
\end{figure}

We report a strong correlation between TeV and X-ray light curves (Swift/XRT, Swift/BAT, MAXI) (Fig. \ref{fig:dcf_all}). We found no significant lag between TeV and X-rays, the combined lag is ($0.26\pm0.46$) days ($1\sigma$). This result is consistent with previously reported lags on more sparse datasets  \citep{2015A&A...576A.126A,2016A&A...593A..91A}. The cross-correlation between GeV and TeV or GeV and X-rays is weak and the lag cannot be identified reliably.

To identify individual flares in X-rays, GeV and TeV bands, we applied the Bayesian Block algorithm \citep{2013ApJ...764..167S} to each light curve and compared the times of the flares, which in this case we define as statistically significant block with amplitude two sigma above the previous block and duration of at least 2 days. The false positive probability was set to 5\% \citep{2013ApJ...764..167S}. All flares based on the Fermi LAT, Swift/XRT, MAXI and FACT light curves are listed in table \ref{tab:flares}. 18 of the TeV flares were detected in the GeV and X-rays, but 4 flares in GeV band were not detected in any other high-energy band. This indicates that particle populations with different spectra are necessary to explain all observations.

\begin{table}[h!]
\centering
\caption{List of gamma ray and X-ray flares sorted by the spectral bands in which they are detected.}
\label{tab:flares}
\begin{tabular}[t]{lcp{7cm}}
Bands & Number & Time ranges, MJD\\
\hline
\hline
TeV only: & 2 & 56689-56692, 57006-57015  \\ \hline
GeV only: & 4 & 56275-56284, 56291-56296, 56368-56376, 57064-57073  \\ \hline
TeV and X-rays: & 11 & 56441-56449, 56696-56700, 57110-57121, 57368-57379, 57385-57390, 57422-57431, 57531-57538, 57728-57753, 57770-57775, 57787-57793, 58103-58113  \\ \hline
TeV, GeV, X-rays: & 18 & 56317-56330, 56369-56383, 56389-56400, 56650-56670, 56751-56755, 56976-57005, 57039-57053, 57065-57070, 57091-57099, 57188-57192, 57432-57449, 57504-57511, 57545-57550, 57756-57769, 57850-57866, 58129-58142, 58162-58167, 58185-58196  \\ \hline
\end{tabular}
\end{table}

The V-band and UV light curves are highly correlated with near to zero lag despite being widely auto-correlated. Both bands are strongly and widely correlated with GeV band with $30-70$ days lag. A similar result is obtained for the radio and GeV correlation, where the correlation is wide with a lag of $40-70$ days at the maximum of the DCF. On the other hand, the radio and optical light curves are not correlated with the TeV light curve, suggesting two separate mechanisms to dominate the production of GeV and TeV gamma rays.

\subsection{GeV to radio response}

\begin{figure}
  \centering
      \includegraphics[width=0.83\columnwidth]{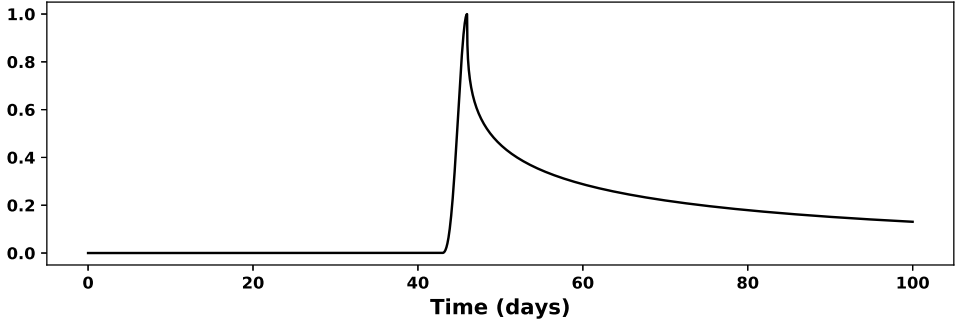}
  \caption{GeV to radio response profile. The $y$ axis is in arbitrary units.}
  \label{fig:radio_profile}
\end{figure}

\begin{figure}
  \centering
      \includegraphics[width=0.83\columnwidth]{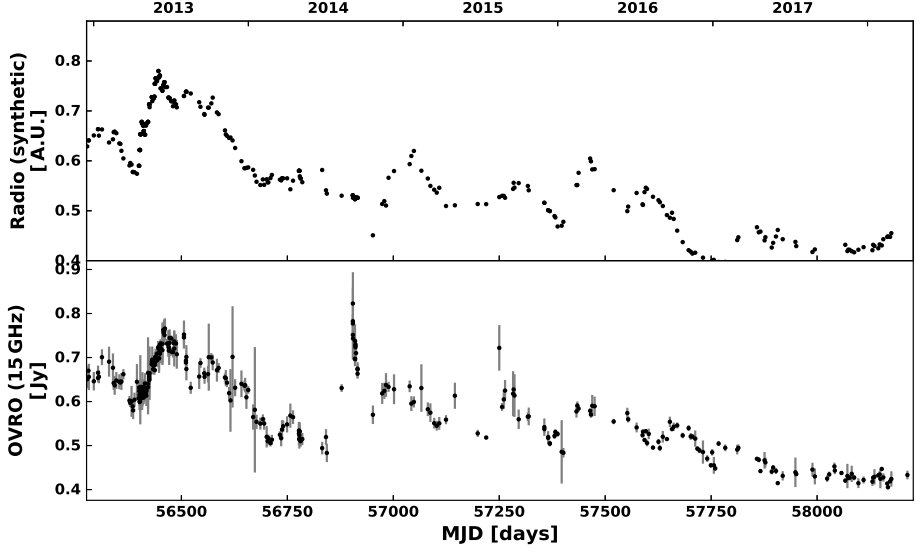}
  \caption{Synthetic radio light curve (top) derived from Fermi LAT light curve and OVRO 15\,GHz radio light curve (bottom). }
  \label{fig:fermi_radio_conv}
\end{figure}

Due to a strong and wide correlation of GeV and radio light curves, we attempted to reconstruct the radio light curve as a convolution of the GeV one with a response profile (see Fig.~\ref{fig:radio_profile}). The original and synthetic radio light curves are shown in Fig.~\ref{fig:fermi_radio_conv}. The response profile is defined by the equations (1) and (2) of section 3.2 of \cite{turler_1999A&A...349...45T}. We find that the profile has $t_{rise}=3$ days, $t_{fall}=7.7$ days, $\rho(\nu)=1.36$, $\phi(\nu)=0.36$. An additional delay $\Delta t = 43$ days was added. Such a delay was also introduced for 3C~273 in \citep{esposito_2015A&A...576A.122E}. We can reproduce the radio light curve ($\chi_{\nu}^2 = 1.2$) except a fast radio flare near MJD 56897.

\section{Results and conclusions\label{sec:conclusions}}

The analysis of 5.5 years long multi-wavelength light curves of Mrk\,421 gives an interesting insight into emission processes of the blazar. We report two main results:

\begin{enumerate}
\item Mrk 421 has the highest variability in X-rays and in the TeV band. 95\% of the short X-ray and TeV flares are coincident. The TeV light curve is strongly correlated with X-rays with no significant lag (($0.26\pm0.46$) days ($1\sigma$)).

\item The radio light curve can be reproduced by convolving the GeV light curve with a fast rise and a slow decay response profile with a delay of $\sim$43 days. 
\end{enumerate}

The fractional variability of Mrk\,421 and the correlated TeV and X-ray emission indicate that the main source of variability is dominated by a synchronous change of the cutoff energies of the low and high-energy components.

The radio emission variability can be reproduced by a delayed response of the GeV variability. This is a strong indication that synchrotron processes dominate the low energy emission component. The radio response, indicating a fast rise after a delay of $\approx$43 days and a slow decay over $\sim$150 days can be interpreted as an emitting region moving outwards and becoming first transparent to gamma rays and later to the radio emission from the radio core. 
The fast rise after a delay may indicate either a discontinuity in the jet direction, in particles density or in magnetic field.

The excellent correlation at zero lag between the TeV and X-ray light curves of Mrk\,421 indicates that these two emissions are driven by the same physical parameter and are consistent with the leptonic emission scenario. This can be driven by variations of the electron maximal energy, or by e.g. the magnetic field that would affect both electrons and protons. The variability of Mrk\,421 is therefore controlled by two independent parameters, the amplitude and the
cut-off energy. On the other hand, the observed variability in the X-rays for protons, according to the hadronic emission scenario, does not match the observations. The proton acceleration times for lepto-hadronic models are also much longer than the time delay between TeV and X-ray light curves yielding that proton-synchrotron emission cannot be responsible for the TeV band.

\noindent\textit{Acknowledgements.} This research has made use of public data from the \textit{OVRO} 40-m telescope \citep{Richards_2011ApJS..194...29R}, the Bok Telescope on Kitt Peak and the 1.54 m Kuiper Telescope on Mt. Bigelow \citep{2009arXiv0912.3621S}, MAXI \citep{2009PASJ...61..999M}, \textit{Fermi} LAT \citep{2009arXiv0912.3621S} and  \textit{Swift} \citep{2004NewAR..48..431G}.

\bibliographystyle{JHEP}
\bibliography{main}

\end{document}